\documentclass[9pt,twocolumn,twoside]{osajnl}

\journal{ol} 

\setboolean{shortarticle}{true}

\usepackage{amssymb}
\usepackage{siunitx}  
\usepackage{graphicx}
\usepackage{hyperref}
\usepackage[utf8]{inputenc}
\usepackage[T1]{fontenc}
\usepackage{color}
\newcommand{\ie}{i.e.~}

\newcommand{\der}{\mathrm{d}}
\newcommand{\unite}[1]{\,\mathrm{#1}}
\newcommand{\eref}[1]{(\ref{#1})}

\title{A high angular resolution interferometric backscatter meter}

\newcommand*{\LAPP}{Laboratoire  d’Annecy  de  Physique  des
  Particules  (LAPP),  Univ. Grenoble  Alpes,
  Université  Savoie  Mont Blanc,
  CNRS/IN2P3,  F-74941  Annecy,  France
}

\author[1,*]{M~Wąs}
\author[1]{E~Polini} 

\affil[1]{\LAPP} 
\affil[*]{michal.was@lapp.in2p3.fr}

\begin{abstract}
  Backscatter limits many interferometric measurements, including gravitational wave detectors, by creating spurious
  interference. We describe an experimental method to directly and
  quantitatively measure the backscatter interference. We derive and
  verify experimentally a relation between backscatter interference,
  beam radius and the scattering sample bidirectional reflectance
  distribution function. We also demonstrate that our method is
  able to measure backscatter from high quality optics for angles as
  low as $500\,\mu\mathrm{rad}$ with a $160\unite{urad}$ angular resolution.
\end{abstract}

\begin{document}

\maketitle

\section{Introduction}

Scattered light is a limitation for many high sensitivity
interferometric measurements.
In particular it is an issue in gravitational wave detectors
such as LIGO and Virgo, which are km scale Fabry-Perot
Michelson interferometers with power and signal recycling \cite{aLIGO,aVirgo}. Scattered light affect
these detectors in two ways: by introducing optical loss that
lowers the optical power and reduces the efficiency of quantum noise
reduction through squeezing \cite{Drori2022}; and by introducing spurious interference
between backscattered light and the main interferometer beam \cite{Vinet97}. The
latter has been a limitation in sensitivity for all interferometric
gravitational wave detectors operated to date and introduces
non-stationary non-Gaussian noise
\cite{GEOHFprogram,aLIGO_O3_scatter,Longo20,Was21}. This is due to
the extreme sensitivity of these detectors where a $10^{-24}$ fraction of
the main beam recombining after a spurious beam path can significantly
affect the sensitivity.

This highlights the importance of understanding backscattered light,
\ie light that leaves the main interferometer beam path, propagates to
a scattering surface and then back-propagates to recombine coherently
with the main interferometer beam. In particular light backscattered at
angles of a few mrad is relevant for  beam expanding
telescopes \cite{Canuel2013}, and at angles between $\sim100\,\mu\mathrm{rad}$ and a few degrees is relevant for core optics baffles
\cite{Vinet97}. Scattering at these small angles is rarely measured
and hard to access by available methods \cite{Drori2022}. 

The angular distribution of light reflected and back-scattered by a surface is usually
characterized by the bidirectional reflectance distribution function
(BRDF). The BRDF is usually measured by directly detecting the scattered light
power using a photo-detector or a camera, and varying the relative
orientation of the sensor and the scattering sample relative to the
probe light beam. These systems can reach a sensitivity of
$10^{-9}\unite{1/srad}$ and an angular resolution of
$300\,\mu\mathrm{rad}$ with up to 15 orders magnitude of dynamic range
\cite{NIST_scatter,Jena_scatter,Labardens2021}. With similar systems scattered light
can also be imaged with lower angular resolution to identify individual point defects contributing
to scattering \cite{MaganaSandoval2012,VanderHyde2015,Kontos2021}. However in most cases
these systems cannot study light that is backscattered directly in the
direction of the incoming beam. Direct backscatter
measurement can easily be limited by the scattering from the
measurement setup reaching only sensitivities of
$10^{-4}\unite{1/srad}$ \cite{Zeidler2019}.

In this paper we present a backscatter measurement that relies on
laser interference and position modulation of the scattering sample
under consideration. A similar approach, but with a more complex
modulation approach and poorer angular resolution, has been used for
scattered light measurements for the future space borne LISA
gravitational wave detector~\cite{Khodnevych2019}. An alternative that
uses a wide spectrum light source~\cite{Khan2021} has achieved a good
angular resolution but two orders of magnitude poorer sensitivity.

\section{Measurement setup}
\label{sec:setup}

\begin{figure}
  \centering
  \includegraphics[width=\columnwidth]{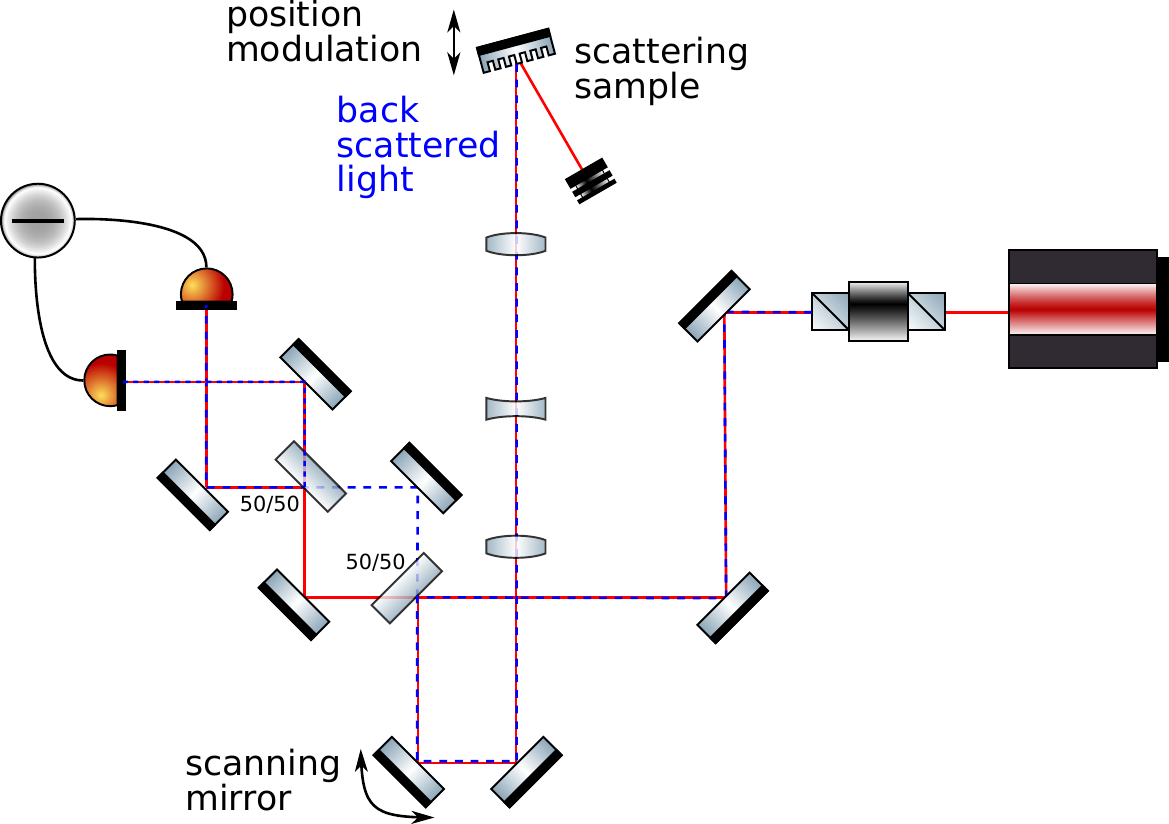}
  \caption{Optical layout of the interferometric scatter meter with
    balanced homodyne detection. The scattering sample is freely
    swinging on a pendulum suspended optical breadboard.}
  \label{fig:layout}
\end{figure}

The measurement optical layout is shown schematically in
figure~\ref{fig:layout}. A single mode S-polarized Nd:YAG laser beam
with $\lambda=1064\unite{nm}$ wavelength and $70\unite{mW}$ power
is split using a 50/50 beam-splitter located at the beam waist. The
beam transmitted by the beam splitter is used as a local oscillator beam for the balanced homodyne
detection (BHD), while the reflected beam is sent to the scattering
sample.  The incident beam is translated across the surface of the scattering
sample using a motorized mirror (Newport AG-M100N). The beam
propagating towards the scattering sample is expanded from a waist of
$300\unite{um}$ to $2.1\unite{mm}$ using a telescope composed of three
lenses. The last two lenses of focal length -100\,mm and 400\,mm compose
a Galilean telescope with magnification $\times 4$. The first lens of 500\,mm
focal length is a relay, which adds a degree of freedom to
simultaneously obtain a collimated beam and to transform the 
angular displacement of the motorized mirror into pure beam translation on the
scattering sample. As a result for lateral beam translations of
$\pm 6\unite{mm}$ the changes of beam tilt on the sample are smaller than  $\pm 30\,\mu\mathrm{rad}$

Backscattered light propagates back through the telescope towards the
first 50/50 beam splitter, where half of the power is sent towards the
BHD and the other half is lost.
The difference in power received by the
two photodectors yields a signal
\begin{equation}
  P_1(t) - P_2(t) = \frac{P_0}{\sqrt{2}} \sqrt{f_\text{sc}}
  \cos\left(2 \pi \frac{\Delta L(t)}{\lambda}\right),
\end{equation}
where $P_0$ is the power provided by the laser, $f_\text{sc}$ is the
fraction of light backscattered by the sample that is mode matched
with the local oscillator beam and $\Delta L(t)$ is the displacement of the
sample. 

This signal can be normalized by the total power received by the two
photodiodes, which yields
\begin{equation}
  A(t) = \frac{P_1(t) - P_2(t)}{P_1(t) + P_2(t)} = \sqrt{2 f_\text{sc}}
  \cos\left(2 \pi \frac{\Delta L(t)}{\lambda}\right).
\end{equation}
For the case where $\Delta L(t)$ fluctuates over many wavelength the
backscattered light fraction is simply obtained as the normalized
signal variance $\textrm{Var}(A(t)) = f_\text{sc}$. Note that this
assumes that the backscattered light is perfectly aligned and mode
matched with the local oscillator beam of the BHD.

\begin{figure}
  \centering
  \includegraphics[width=\columnwidth]{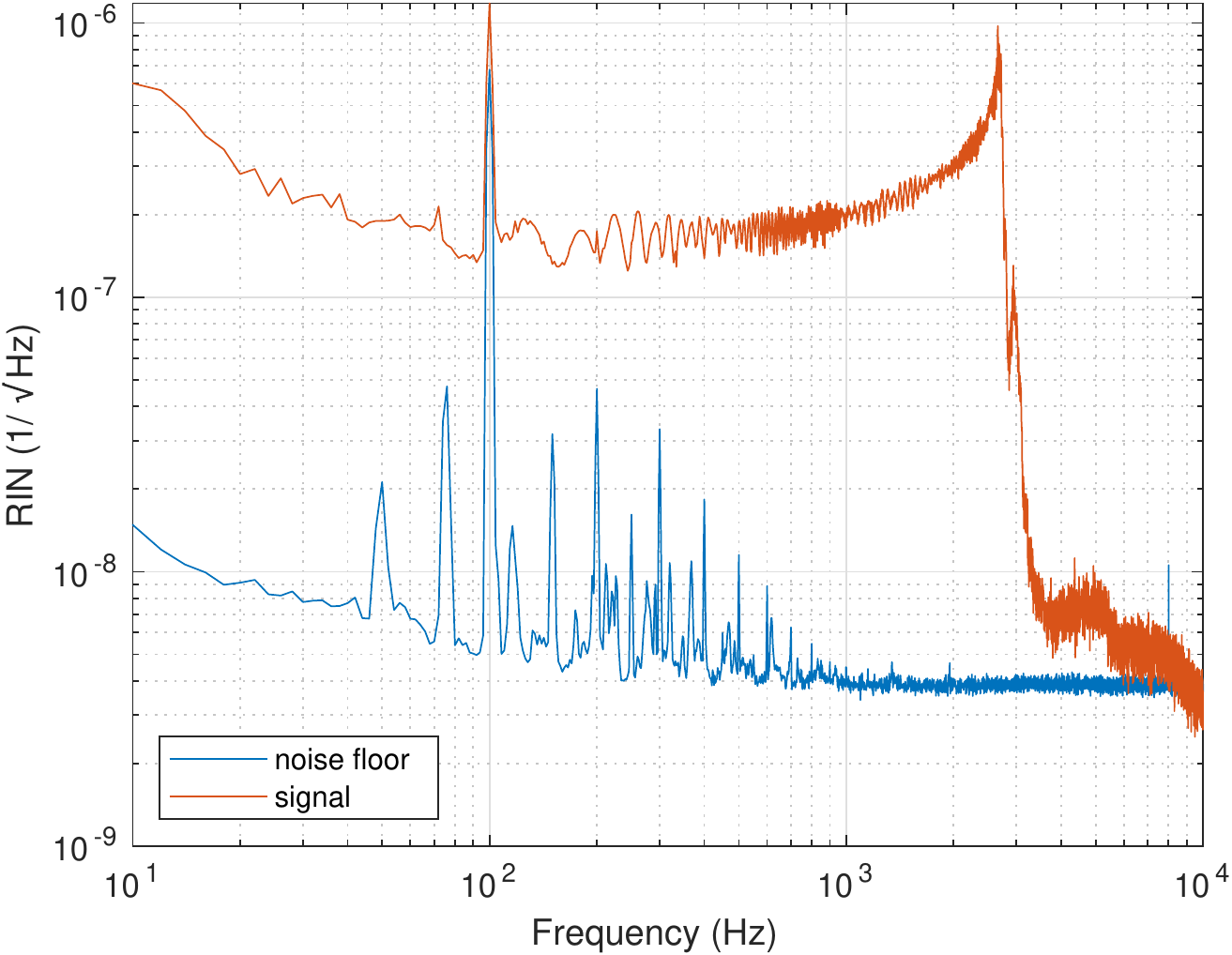}
  \caption{ Example amplitude spectral density of the normalized BHD
    signal $A$ with a scattering sample of
    $f_\text{sc}=3 \times 10^{-10}$ is shown in red, and the average
    measurement noise floor without any sample in blue.}
  \label{fig:PSD_example}
\end{figure}

To achieve a large position modulation the sample is placed on a breadboard
suspended on four metal wires at $\sim45$ degrees angle to the
vertical, similarly to a Newton's cradle. This allows a free pendulum
motion along the beam axis, while restricting the motion in the
orthogonal plane. After a gentle touch the free motion of that suspension creates
a decaying sinusoidal motion with an amplitude of a few hundred
wavelengths and quality factor $\sim1000$.  The motion Doppler shifts
the backscattered light frequency and create a quasi sinusoidal signal with
a time dependent frequency
$f(t) = \frac{2\left|\dot{x}(t)\right|}{\lambda}$ proportional to the
instantaneous bench speed. This results in a signal with a power
spectrum spreading between $0\unite{Hz}$ and a few kHz, with a peak
power just below the cut-off frequency as shown in
figure~\ref{fig:PSD_example}. As the sample is the only object with a
large motion, the up-conversion clearly identifies light that is back
reflected or backscattered by the sample.

Above 500\,Hz the noise floor is shot noise limited, while below there are
significant contributions of the mechanical resonances of the various
optical components used in the measurement and 50\,Hz mains AC current
harmonics. To measure $f_\text{sc}$
without being affected by these noise sources we
integrate the power spectrum density of $A$ only between 500\,Hz and
6\,kHz, which yields a variance measurement with negligible bias as
long as the peak emission is between 2\,kHz and 5\,kHz.

This measurement is affected by speckle, that is by the particular
random realization of the light that is backscattered. Hence the
measured quantity is an exponentially distributed random
variable. This random effect can be averaged by scanning the
scattering sample in translation, with each translation by a beam
radius yielding an independent measurement~\cite{Khodnevych2019}.
We scan the target surface with aid of the motorized
mirror using 49 dedicated points on the surface spanning over an area of 7x7 beam
radii. This yields a measurement of the average
scattering with $\frac{1}{\sqrt{49}}\simeq14\%$ statistical errors. This inherently assumes that
the target scattering is uniform over the scanned surface. This is a
good assumption for rough surfaces however for higher quality optics
localized scratches or dust particles can yield high scattering for
some pointings. To mitigated this we use a median of the 49
measurements corrected by a factor $\log 2$ to obtain a robust
measurement of the exponential distribution parameter.

\section{Relation between $f_\text{sc}$ and BRDF}
\label{sec:theory}

In the previous section we have described how the scattered light
fraction $f_\text{sc}$ can be measured using a BHD. This is the
quantity of interest when evaluating the impact of scattered light on
gravitational wave detectors or other interferometric
measurements. However, scattered light of a given object is usually
characterized using its BRDF. In this section we will derive a simple
analytical relation between these two quantities.

A normalized gaussian beam with radius $w(z) = w_0 \sqrt{1 +
  \frac{\lambda^2 (z-z_0)^2}{\pi^2 \omega_0^4}}$ and wavefront radius of
curvature $R(z)$ has the following form in cylindrical coordinates
\begin{equation}
  E_\text{beam}(r, z) = \sqrt{\frac{2}{\pi w(z)^2}} \exp\left[ - \left(\frac{1}{w(z)^2}
    + \frac{\pi i}{\lambda R(z)}\right) r^2\right].
\end{equation}
We mean by normalized that
$\int_0^\infty 2\pi r |E_\text{beam}|^2 \der r= 1$. For a point scatterer the
light close to the optical axis at large distance $z$ can be
approximated by a normalized spherical wave
\begin{equation}
  E_\text{sc}(r, z) = \frac{1}{\sqrt{2 \pi z^2}} \exp\left[ - \frac{\pi i}{\lambda z} r^2\right].
\end{equation}

The overlap integral of these two fields will yield the interference between a
Gaussian beam and a spherical wave, \ie a perfect scatterer.
\begin{eqnarray}
  \text{Overlap} &= \left|\int_0^\infty 2 \pi r E_\text{beam} E^*_\text{sc}
                   \der r \right|^2\\
  &= \left|\int_0^\infty \frac{2 r \der r}{w z} \exp\left[ - \left(\frac{1}{w^2}
    + \frac{\pi i}{\lambda R} - \frac{\pi i}{\lambda z}\right)
    r^2\right]\right|^2 \\
  &= \frac{1}{\frac{z^2}{w^2} +
    \frac{\pi^2}{\lambda^2}\left(\frac{1}{R} - \frac{1}{z}\right) w z}.
\end{eqnarray}
This expression can be further approximated by evaluating it in the
Gaussian beam far field, \ie assuming that $z$ is many Rayleigh ranges
from the beam waist of radius $w_0$ at position $z_0$.
\begin{equation}\label{eq:overlap}
  \text{Overlap} \simeq \frac{1}{\frac{z^2}{w^2} +
                   \frac{\pi^2}{\lambda^2}\frac{z_0^2 w^2}{z^2}}
  \simeq\frac{1}{\frac{\pi^2 w_0^2}{\lambda^2} +
     \frac{z_0^2}{w_0^2}}
    \simeq \frac{\lambda^2}{\pi^2 w(0)^2},
\end{equation}
where $w(0)$ is the Gaussian beam radius at the position of the
spherical wave emission.

Hence the overlap between a Gaussian beam and backscattered light
depends only on the size of the beam at the scattering
surface. Although this was derived for a point scattering source
centered on the beam axis, it remains true if the scattering is due to a large
number of uniformly distributed point sources or due to random
roughness of the surface. We will not derive this more general
relation, but verify it experimentally in section~\ref{sec:results}.

A normalized spherical wave is equivalent to a BRDF of $\frac{1}{\pi
  \cos(\theta)}$, where $\theta$ is the incidence angle on the scattering object.
Hence for a scattering object the backscattered light fraction is
simply obtained by replacing the spherical wave BRDF by the BRDF of
the sample object 
\begin{equation}\label{eq:BRDF2fsc}
  f_\text{sc} = \text{Overlap} \times \pi \text{BRDF}(\theta)  \cos \theta  =
  \text{BRDF}(\theta) \frac{\lambda^2 \cos \theta}{\pi w(0)^2}.
\end{equation}

\section{Results}
\label{sec:results}

\begin{figure}
  \centering
  \includegraphics[width=\columnwidth]{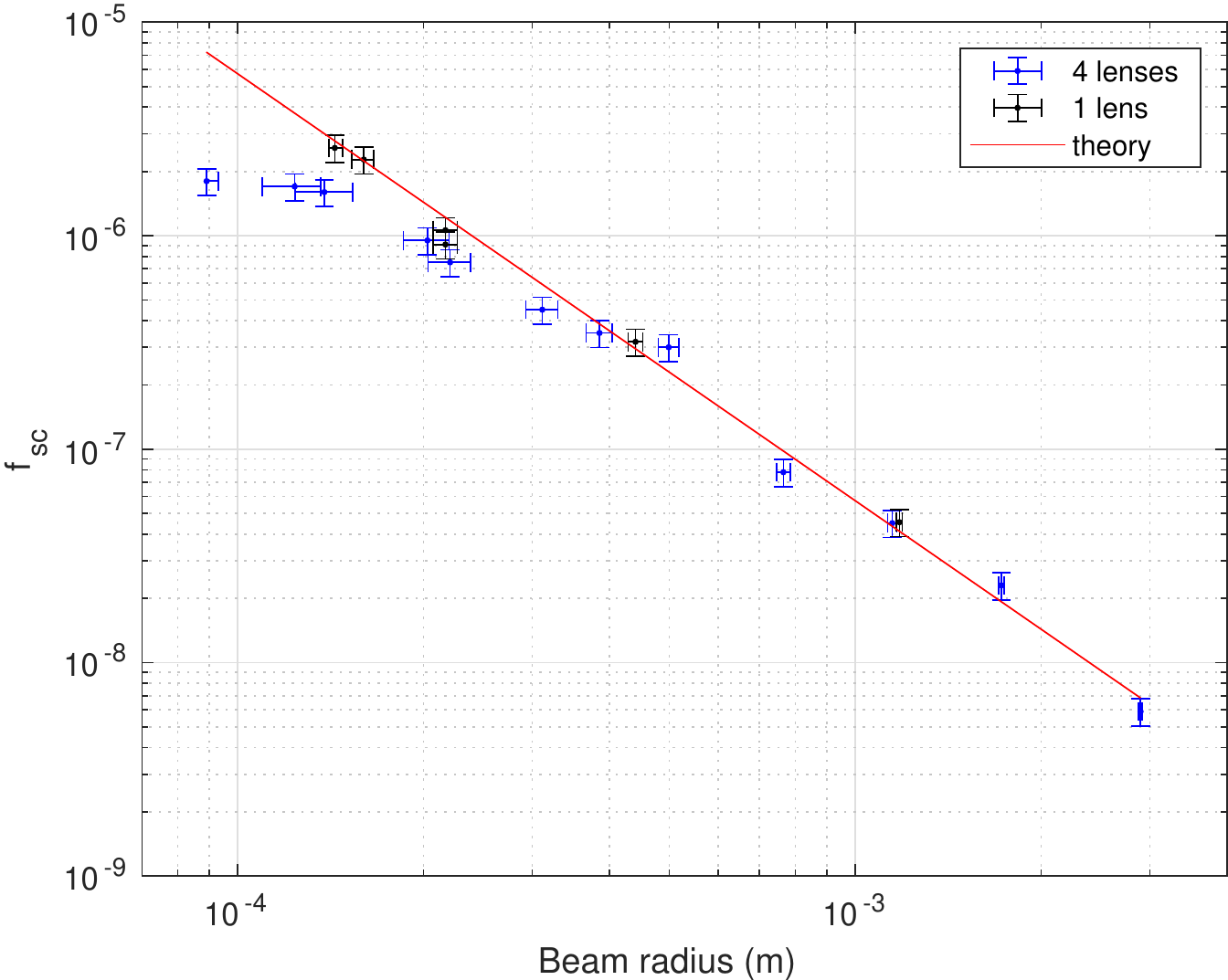}
  \caption{Measured backscatter fraction $f_\text{sc}$ as function of
    beam size for the four lens configuration (blue) and the single
    lens configuration (black), compared to the expectation given by
    equation~\eref{eq:BRDF2fsc} in red. Vertical error bars correspond
    to the statistical error of the measurement due to speckle
    averaging, while the horizontal error bars are the target
    positioning errors of $\pm 5\unite{mm}$ re-stated as errors on the
    beam radius on the target.}
  \label{fig:fsc_vs_distance}
\end{figure}

We have performed measurements using the setup described in section~\ref{sec:setup}.
The alignment of the backscattered field with the local oscillator
beam has been verified using a flat
mirror instead of the scattering target shown in
figure~\ref{fig:layout}. The directly back-reflected beam has
interference visibility of 88\%, while the maximum that
is attainable is 94\% given that half of the retro-reflected light is
lost and sent back towards the laser. The measured backscatter
fraction below is corrected for this interference visibility loss.

\subsection{Backscatter as a function of beam radius}

To verify the relation between backscatter fraction $f_\text{sc}$ and
 BRDF  we have measured
the backscattering from a PTFE target (integrating sphere plug) as a
function of beam size. PTFE is a near perfect volume Lambertian
diffuser that scatters light in both polarizations with ideally a
$\text{BRDF}=\frac{1}{2 \pi}$ \cite{Bhandari2011}. The beam radius
was measured with a CCD camera and fitted with a Gaussian
beam independently in the horizontal and vertical direction. The
root-mean-square of the horizontal and vertical radius is used as the
beam radius in the results below.

The beam at the output of the telescope described in
section~\ref{sec:setup} was converged with a $500\unite{mm}$ focal
length lens to a waist of $85\unite{um}$ and the target placed at
different locations to vary the beam size by more than one order of
magnitude. The measured backscatter fraction is shown in
figure~\ref{fig:fsc_vs_distance} and matches well the theoretical
expectation of equation~\ref{eq:BRDF2fsc} for beam radii above
$200\unite{um}$, but it is a factor 2-3 smaller close to the beam
waist. This may be due to aberration in the beam converged using a
single plano-convex lens. To verify this instead of four lenses we
used a single $175\unite{mm}$ focal length lens and a shorter optical
path length. This configuration shows an excellent match at all radii down to the waist of
$145\unite{um}$.

These results confirm the relation given by
equation~\eref{eq:BRDF2fsc}.

\subsection{Backscatter as a function of incidence angle}

\begin{figure}
  \centering
  \includegraphics[width=\columnwidth]{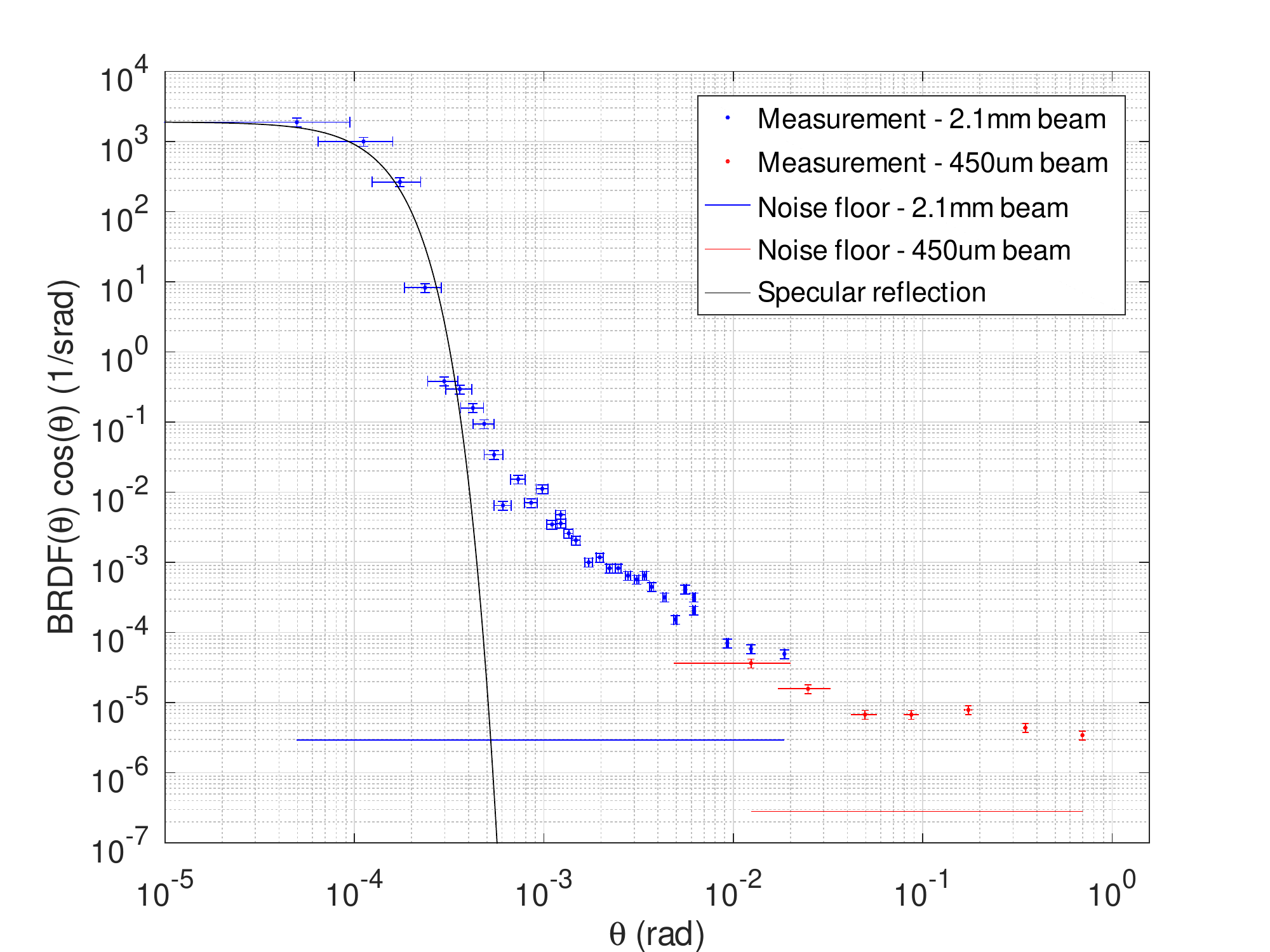}
  \caption{Measured BRDF using a 2.1\,mm radius beam (blue points) and
    $450\unite{um}$ radius beam (red points) as a function of
    incidence angle $\theta$. The horizontal blue and
    red solid lines represents the corresponding measurement noise
    floor observed without a sample. The black solid line show the
    expected interference from the window 700\,ppm specular reflection.}
  \label{fig:brdf_vs_angle}
\end{figure}

The backscatter as a function of incidence angle was measured for a
multi-band ($512\unite{nm}/800\unite{nm}/1064\unite{nm}$) anti-reflective coated vacuum
window. The coating has been measured to have a 700\,ppm reflectivity
at $1064\unite{nm}$ for 0 degree incidence angle. The 0.5 degree wedge
of the window was placed vertically, and the incidence angle changed
horizontally to measure at small angles the backscatter from only one
of the two surfaces. The measured $f_\text{sc}$ is converted into BRDF
using equation~\eref{eq:BRDF2fsc}.

The measurement result is shown in
figure~\ref{fig:brdf_vs_angle}. Below $500\,\mu\mathrm{rad}$ the tails
of the Gaussian beam specular reflection are dominant as the beam
provided by the telescope has a $160\,\mu\mathrm{rad}$ divergence. The
angular errors are due to the quadratic sum of $\pm 30\,\mu\mathrm{rad}$
beam incidence angle variation during the sample surface scan and the
$\pm 30\,\mu\mathrm{rad}$ positioning error provided by the window
kinematic mount micrometer (Newport HR-13). Note that within 1\,m of
the beam waist the Gaussian beam wavefront curvature is essentially
flat, inducing negligible angular changes of less than
$6\,\mu\mathrm{rad}$ across the beam radius.

At angles larger than 10\,mrad the optical configuration was changed
by removing the diverging lens from the optical path to obtain a
smaller beam radius of $450\unite{um}$ and a $\sim15$ times larger signal
above the measurement noise floor. However this introduces a much larger
$\pm 8\unite{mrad}$ angular error as in that case the motorized mirror
motion no
longer preserves the incidence angle on the sample.

For angles between 1\,mrad and 10\,mrad the BRDF is proportional to
$\frac{1}{\theta^2}$, which is typical for polished
optics \cite{Canuel2013,Drori2022}. While at angles larger than 50\,mrad the BRDF becomes
approximately independent of the incidence angle, this transition occurs at
a smaller angle compared to uncoated or high reflective coated optics
because the $\propto \frac{1}{\theta^2}$ term is strongly reduced by
the anti-reflective coating \cite{Amra1986, Soriano2019}.

\section{Conclusion}
\label{sec:conclusion}

We have proposed and implemented an interferometric measurement of
backscattered light that uses relatively simple and readily available
components. In a simplified case we have derived the relation between
the backscattered light fraction, the Gaussian beam radius and the
sample BRDF. This relation was verified experimentally using the
proposed setup.

The measurement setup achieved a $160\,\mu\mathrm{rad}$ angular
resolution limited by beam divergence and measures backscatter for incidence angles larger than
$500\,\mu\mathrm{rad}$. In particular it is able to measure backscatter
at angles of a few mrad that are relevant for beam expanding telescope
lenses in gravitational wave detectors. This allows to measure the
scattering of coated optics, which depends on the
surface roughness and defects, but also on the scattering reduction
due to the anti-reflective coating that depends on how well the
coating follows the bare optic surface.

This method can be further expanded  to increase the telescope
magnification by a factor 10 to reach a beam radius of $20\unite{mm}$.
This should allow to measure backscatter for angles larger than
$50\,\mu\mathrm{rad}$ with $17\,\mu\mathrm{rad}$ resolution, and cover angles of $\sim100\,\mu\mathrm{rad}$
to measure the most critical scattering from gravitational wave
detectors core optics. As translating a large beam by several waist
becomes increasingly impractical, the speckle can be instead averaged
by changing the beam tilt by several beam divergence angles, for instance in a circle at fixed incidence with
respect to the sample surface.

\begin{backmatter}
  \bmsection{Acknowledgements} The Virgo document number of this paper
  is VIR-0111A-22.
  \bmsection{Data Availability Statement} Data underlying the results
  presented in this paper are not publicly available at this time but
  may be obtained from the authors upon reasonable request.
  \bmsection{Disclosures} The authors declare no conflicts of interest.
\end{backmatter}

\end{document}